\documentclass[a4paper,11pt,preprint,showpacs,superscriptaddress,preprintnumbers,prd]{revtex4}
\usepackage{graphicx,amsmath,dcolumn,bm}
\usepackage{hyperref}
\usepackage{epstopdf}
\begin{document}
\title{\Large{\bf{Bjorken variable and scale dependence of quark transport coefficient in Drell-Yan process for proton incident on nucleus}}}

\author{Tian-Xing Bai}
\email[E-mail:]{txbai@hebtu.edu.cn}
\affiliation{College of Physics and Hebei Advanced Thin Film Laboratory, Hebei Normal University, Shijiazhuang 050024, P.R.China}

\author{Chun-Gui Duan}
\email[E-mail:]{duancg@hebtu.edu.cn}
\affiliation{College of Physics and Hebei Advanced Thin Film Laboratory, Hebei Normal University, Shijiazhuang 050024, P.R.China}

\begin{abstract}

By means of the nuclear parton distributions determined only with lepton-nuclear deep inelastic scattering experimental data and the analytic parameterization of quenching weight based on BDMPS formalism, a phenomenological analysis of the nuclear Drell-Yan differential cross section ratio as a function of Feynman variable is performed from Fermilab E906 and E866 experimental data. With the nuclear geometry effect on nuclear Drell-Yan process and the quark transport coefficient as a constant, our predictions are in good agreement with the experimental measurements. It is found that nuclear geometry effect has a significant impact on the quark transport coefficient in cold nuclear matter. It is necessary to consider the detailed nuclear geometry in studying the nuclear Drell-Yan process. Our calculated results reveal that the difference in the values of quark transport coefficient exists from E906 and E866 experiments. However, confirming the conclusion that the quark transport coefficient depends on the target-quark momentum fraction, still need more accurate experimental data on the Drell-Yan differential cross section ratio in the future. Three models are proposed and discussed for the quark transport coefficient as a function of the measurable kinematic variables. The quark transport coefficient is determined as a function of the Bjorken variable $x_2$ and scale $Q^2$.

\vskip 0.1cm



\end{abstract}

\maketitle

\newpage
\vskip 0.5cm

\section{Introduction}

Studying deeply how partons propagate in a nucleus can help us to reveal precious information about the transport and thermodynamical properties of the quark-gluon plasma(QGP)[1-3] which is formed in ultra-relativistic heavy-ion collisions. As the scattered partons traverse through and interact with the QGP, they lose some of their energy, which is often known as jet quenching observed at the BNL Relativistic Heavy Ion Collider (RHIC)[4, 5] and the CERN Large Hadron Collider (LHC)[6-8]. However, the mechanisms by which these partons experience the full evolution of the medium and interact with the medium constituents, are still not fully understood[9-12].

As for the Drell-Yan reactions[13] on nuclei, the bound nucleons in target nuclei play the role of very nearby detectors for the propagating partons. In contrast to heavy ion collisions, the hard scattering is localized in space. In addition, the properties of target nuclei in which the parton propagates, i.e. cold nuclear matter, is well known. Furthermore, the Drell-Yan reactions on nuclei is an especially clean tool, since the produced dilepton, which has no final state interactions, carries undisturbed information on the incident parton initiating the reaction. Therefore, the Drell-Yan reaction in hadron-nucleus collisions provide an excellent place to study how partons propagate in a nucleus[14-18, 27-31].

A quark passing through a nuclear medium experiences the multiple interactions with bound nucleons in target nuclei. Quark multiple scattering is responsible for its energy loss and transverse momentum broadening. The transverse momentum broadening in the Drell-Yan reaction is defined as
\begin{equation}
\Delta\langle P_T^2\rangle=\langle P_T^2\rangle_{\rm hA}-\langle P_T^2\rangle_{\rm hN},
\end{equation}
where $\langle P_T^2\rangle_{\rm hA}$ and $\langle P_T^2\rangle_{\rm hN}$ are respectively the mean transverse momentum squared of the dilepton produced in hadron-nucleus (hA) and hadron-nucleon(hN) collisions. Baier et al.[19-21] (BDMPS hereafter) studied the medium-induced transverse momentum broadening and induced gluon radiation spectrum of a high energy quark traversing a nucleus. Multiple scattering of the high energy quark in the nucleus is treated in the Glauber approximation. BDMPS showed the connection between the radiative energy loss of the quark per unit length and the transverse momentum broadening. Moreover, the transverse momentum broadening for the quark is given by
\begin{equation}
\Delta\langle P_T^2\rangle=\hat{q}L ,
\end{equation}
with the length $L$ crossed by the incident parton in a target nucleus. The quark transport coefficient is determined by
\begin{equation}
\hat{q}=\frac{4\pi^{2}\alpha_{\rm s}(Q^{2}_{\rm G})C_{F}}{N^{2}_{c}-1}\rho x_{\rm G}G(x_{\rm G},Q^{2}_{\rm G}),
\end{equation}
where $\rho$ is the nuclear matter density, $C_{F}$ is the quark colour factor, the number of colors $N_{c}=3$, the virtuality $Q^{2}_{\rm G} = \hat{q}L$. The strong coupling constant $\alpha_{\rm s}$ and the gluon distribution function $G(x_{\rm G},Q^{2}_{\rm G})$ depend on the scale $\hat{q}L$. The relative calculation indicates that the Bjorken variable $x_{\rm G}\ll1$ in the gluon distribution function.

CERN experiment NA10[22] made the first measurements of the nuclear dependence of the transverse momentum distribution of massive muon pairs produced in the Drell-Yan process. The cross sections for high-mass dimuon production by incident negative pions (140 and 286 GeV) were compared off tungsten and deuterium. They found that the transverse momentum broadening amounts to $0.15 \pm 0.03(\rm stat.) \pm 0.03 (\rm syst.) $ GeV$^2$ at 286 GeV. Fermilab experiment E772[23] extended the nuclear dependence measurements on the transverse momentum broadening of dimuon production by an 800 GeV proton beam incident on $^2$H, C, Ca, Fe, and W targets. They presented that the transverse momentum broadening was respectively $0.0 \pm 0.015 $ GeV$^2$ for C nucleus, $0.046 \pm 0.011 $ GeV$^2$ for Ca nucleus, $0.048 \pm 0.012 $ GeV$^2$ for Fe nucleus and $0.113 \pm 0.016 $ GeV$^2$ for W nucleus. Obviously, the experimental data from NA10 and E772 show that $\hat{q}L<1$ GeV$^{2}$. Therefore, the virtuality $Q^{2}_{\rm G} $ is less than 1 GeV$^2$ in the gluon distribution function from the quark transport coefficient.

As for the quark transport coefficient $\hat{q}$, the Bjorken variable $x_{\rm G}$ estimated in theory is an unmeasurable kinematic variable in experiment. The unmeasurable $x_{\rm G}$ prevents us from knowing the detail of the gluon distribution function $G(x_{\rm G}, Q^{2}_{\rm G})$. The undiscovered detail on the quark transport coefficient limits the predictive power of the BDMPS theory. Therefore, in order to understand the property of the quark transport coefficient, one would like to know whether the connection can be made between the unmeasurable $x_{\rm G}$ and the measurable kinematic variable in Drell-Yan process. The purpose of this paper is to establish the relation between quark transport coefficient $\hat{q}$ and the measurable kinematic variable in Drell-Yan process.

In fact, the scale dependent of jet transport parameter has been derived theoretically in 2014[24]. Recently, by using the experimental data on transverse momentum broadening from semi-inclusive electron-nucleus deep inelastic scattering, Drell-Yan dilepton and heavy quarkonium production in proton-nucleus collisions, as well as the nuclear structure functions in deep inelastic scattering, the kinematic and scale dependence of jet transport coefficient in cold nuclear matter has been determined in the framework of the generalized QCD factorization formalism[25]. By means of the world data on nuclear transverse momentum broadening from Drell-Yan and quarkonium production for pion and proton incident on nucleus, the transport coefficient parameterization as a function of Bjorken variable has been extracted[16]. In order to study the underlying structure of the quark-gluon plasma (QGP), the jet transport coefficient was suggested to be expressed in terms of a parton distribution function with energy and scale dependence[26].

In our preceding articles[27-31], we have investigated the energy loss effect on the nuclear Drell-Yan reactions. In the recent work[32], we take advantage of the analytic parameterization of quenching weight based on BDMPS formalism[33] with the target nuclear geometry effect. The leading-order computations for hadron multiplicity ratio are carried out with comparison to the HERMES positively charged pions production data[34] on the quarks hadronization occurring outside the nucleus. Four models are proposed and discussed on the quark transport coefficient. The quark transport coefficient is determined as a function of the Bjorken variable $x$ and scale $Q^2$ in deep inelastic scattering. We found that the trend of quark transport coefficient in respect of the Bjorken variable $x$ and scale $Q^2$ is qualitatively in partial agreement with HERMES experimental data on transverse momentum broadening[35]. The constant factor in the quark transport coefficient is almost twice the quark transport coefficient as a constant.

The main goal of the present work is to explore the Bjorken variable and scale dependence of quark transport coefficient by a fit of the nuclear Drell-Yan differential cross section ratios from E906[36] and E866[37] Collaborations at Fermilab. The BDMPS quenching weight[33] will be employed with simultaneously considering the nuclear geometry effect and nuclear effects on parton distribution. It is hoped to provide a good understanding of the quark transport coefficient in a cold nuclear medium from the available data, and to facilitate the theoretical research on the parton propagation mechanism in nuclear matter.

The paper is organized as follows. In Section II we give the brief formalism for the differential cross section in nuclear Drell-Yan process, the obtained results and discussion are presented in Section III. Finally, we summarize our findings in Section IV.

\section{Dilepton production differential cross section on nuclear targets}

According to the parton model, the produced lepton pair can be naively explained as the annihilation of quark-antiquark of the same flavor from the colliding hadrons in the lowest approximation, and this is what Drell and Yan suggested[13]. In the framework of the collinear factorization, the proton-nucleus Drell-Yan differential cross section in $x_{\rm F}$ and $M$ is given by
\begin{equation}
\frac{d^2\sigma}{dMdx_{\rm F}}=\frac{8\pi\alpha_{\rm em}^2}{9Ms}\frac{1}{x_1+x_2}H(x_1,x_2,Q^2),
\end{equation}
with
\begin{equation}
H(x_1,x_2,Q^2)=\sum_{f}e^2_f[q^{\rm p}_f(x_1,Q^2)\bar{q}^{\rm A}_f(x_2,Q^2)+\bar{q}^{\rm p}_f(x_1,Q^2)q^{\rm A}_f(x_2,Q^2)],
\end{equation}
where $\alpha_{\rm em}$ is the fine structure coupling constant, $\sqrt{s}$ is the center of mass energy of the hadronic collision, the factorization scale $Q^2=M^2$, $M$ is the invariant mass of a lepton pair, $e_f$ is the charge of the quark, the sum is carried out over the light flavor $f={\rm u,d,s}$, $q^{\rm p(A)}_{f}(x,Q^2)$ and ${\bar q}^{\rm p(A)}_{f}(x,Q^2)$ are the quark and antiquark distributions in the proton (nucleon in the nucleus A), $x_1$ and $x_2=M^2/(x_1s)$ are the momentum fraction of the partons in the beam and target respectively. As for the Feynman variable of the lepton pair, $x_{\rm F}=x_1-x_2$, one can easily find
\begin{equation}
x_1=\frac{1}{2}(\sqrt{x_{\rm F}^2+4\frac{M^2}{s}}+x_{\rm F}),\quad x_2=\frac{1}{2}(\sqrt{x_{\rm F}^2+4\frac{M^2}{s}}-x_{\rm F}).
\end{equation}
In the Drell-Yan reactions on nuclei, the incident quark propagating in a target nucleus suffers multiple scattering and radiates soft gluons. This multiple scattering effect results in the energy loss of an incoming quark. At the time of hard interaction, the quark momentum fraction is reduced from $x_1+\Delta x_1$ to $x_1$ with $\Delta x_1 =\Delta E/E_{\rm p}$. $\Delta E$ and $E_{\rm p}$ is respectively the quark energy loss in the nuclear medium and the projectile hadron energy. Consequently, with adding the quark energy loss in the nucleus,
\begin{equation}
H(x_1,x_2,Q^2)=\int_{0}^{(1-x_1)E_{\rm p}}d(\Delta E) P(\Delta E,\omega_{c},L)H(x_1+\Delta x_1, x_2,Q^2).
\end{equation}
$P(\Delta E,\omega_{c},L)$, the so-called quenching weight, is the probability that the radiated gluons carry altogether a given energy $\Delta E$ for an incident quark. The characteristic gluon frequency $\omega_{c}=(1/2)\hat{q}L^{2}$ with the path length $L$ traversed by the incoming quark or antiquark.

For simplicity, we consider the case of the uniform hard-sphere nucleus of mass number $A$ and radius $R_A=r_0{A^{1/3}}$ with $r_0 = 1.12$ fm[38]. The average path length of the incident quark in the nucleus is given by $L=(3/4)R_A$. With referring to Ref.[39], in view of the nuclear geometry effect, the colored incoming quark interacting at $y$ the coordinate along the direction of the incident quark will travel the path length $L=\sqrt{R_{A}^{2}-b^{2}}+y$ with $\vec{b}$ its impact parameter. After taking both nuclear geometry and parton energy loss effects into account,
\begin{equation}
H(x_1,x_2,Q^2)= \int d^{2}bdy\rho_{A}(\vec{b},y)\int_{0}^{(1-x_1)E_{\rm p}}d(\Delta E) P(\Delta E,\omega_{c},L)H(x_1+\Delta x_1, x_2,Q^2).
\end{equation}
In above equation, $\rho_A(\sqrt{b^2+y^2})=(\rho_0/A)\Theta(R_A-\sqrt{b^2+y^2})$ with $\rho_0$ the nuclear density.

Based on the above formalism, only the transport coefficient $\hat{q}$ remains to be determined.

\section{Results and discussion}

In order to study the Bjorken variable and scale dependence of the quark transport coefficient $\hat{q}$ in the cold nuclear medium, we employ the experimental data on nuclear Drell-Yan differential cross section ratio from E906[36] and E866[37] Collaborations at Fermilab. E866 Collaboration reported their precise measurement of the ratios of Drell-Yan cross section per nucleon for an 800 GeV proton beam incident on Be, Fe, and W targets. The kinematic coverage extended over the ranges $0.01<x_2<0.12$, $0.21<x_1<0.95$ and $0.13<x_{\rm F}<0.93$ with $4.0<M<8.4$ GeV. E906 Collaboration used an 120 GeV proton beam incident on C, Fe and W to study Drell-Yan reaction dimuons. The experiment measured the ratio on Drell-Yan yield of Fe/C and W/C as a function of $x_{\rm F}$. The kinematic range is $0.1< x_2 < 0.3$ with $4.5< M < 5.5$ GeV. Our analysis has 28 data points in total, includes 12 data points from E906 experiment, and 16 data points from E866 measurement on the ratios of the cross section per nucleon versus $x_{\rm F}$.

To compare with the experimental data on nuclear Drell-Yan differential cross section ratio, we calculate in perturbative QCD leading order the Drell-Yan cross section ratio on two different nuclear targets bombarded by proton,
\begin{equation}
R_{\rm A_1/A_2}(x_{\rm F})=\int\frac{d^2\sigma^{\rm pA_1}}{dMdx_{\rm F}}dM\Bigg/\int\frac{d^2\sigma^{\rm pA_2}}{dMdx_{\rm F}}dM.
\end{equation}

\begin{figure}
\centering
\includegraphics*[width=15cm,height=10.7cm]{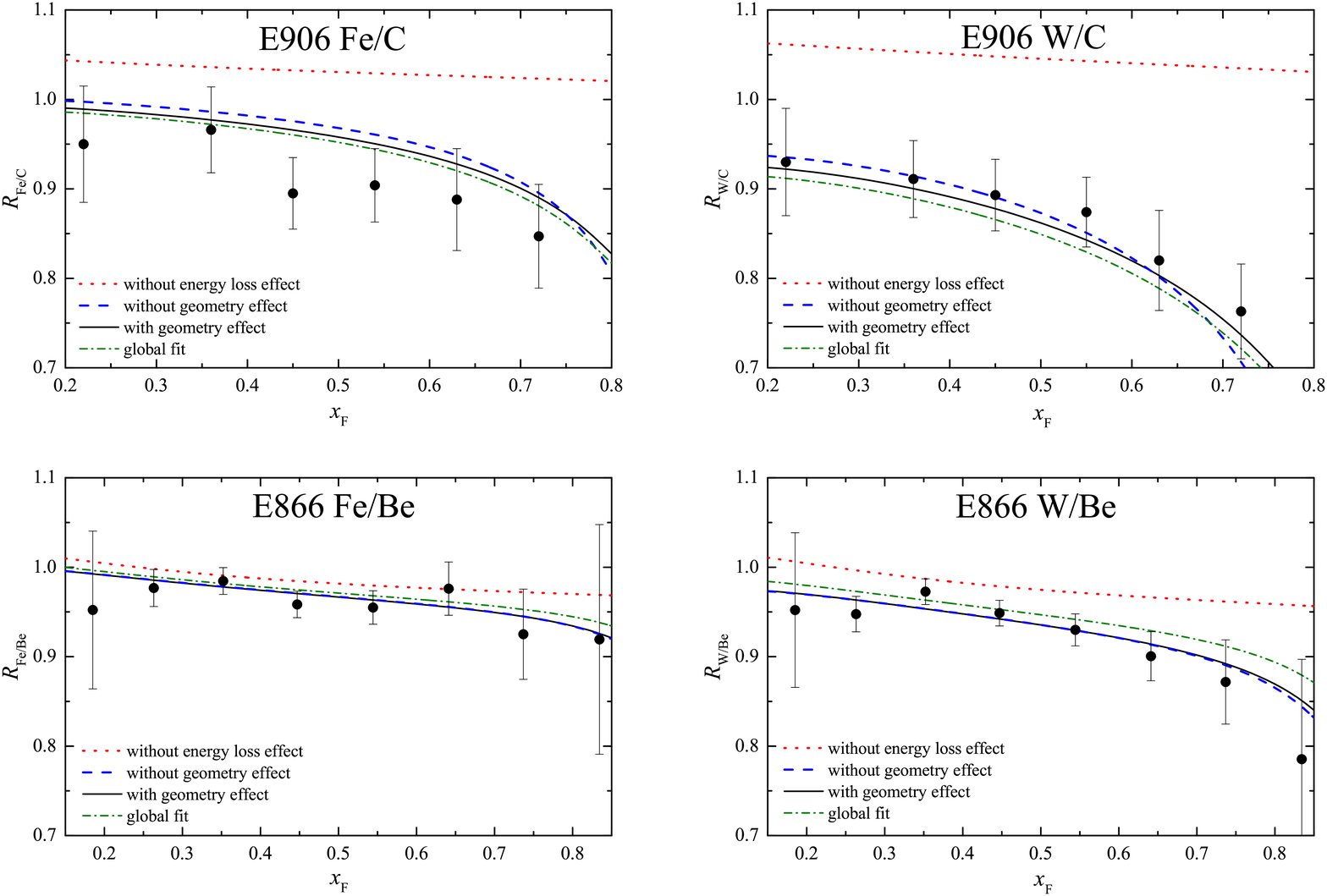}
\vspace{-0.1cm}
\caption{The calculated Drell-Yan cross section ratios $R_{\rm A_1/A_2}$ as a function of $x_{\rm F}$. As for the fitting respectively E906 and E866 experimental data, the red dotted lines are the calculation without energy loss effect. The black solid (blue dashed) lines correspond to the predictions with (without) target nuclear geometry effect. As for the fitting totally E906 and E866 experimental data, the green dot-dashed lines are the computed results with target nuclear geometry effect. The experimental datas are taken from E906 and E866 Collaboration at Fermilab and shown with the total uncertainty (statistical plus systematic, added quadratically).}
\end{figure}

In order to explore the nuclear effects from parton distribution functions on the Drell-Yan differential cross section ratio, the nuclear parton distribution functions given by Hirai, et al.(HKM hereafter)[40] are used in our calculations without the consideration of quark energy loss effect. It should be noted that HKM nuclear parton distributions were determined only by means of the existing experimental data on nuclear structure functions without including the proton-nucleus Drell-Yan process. The $\chi^2$ analysis are performed respectively to the E906 and E866 experimental data. Our numeric result gives $\chi^2$ per number of degrees of freedom $\chi^2/{\rm ndf} =10.960$ for E906 and $\chi^2/{\rm ndf} =2.281$ for E866 experimental data. The calculated results are compared with the E906 and E866 experimental data on Drell-Yan differential cross-section ratios in Fig.1(red dotted lines). Obviously, the results with only considering the nuclear effects of parton distribution can not explain the employed experimental data.

In the following part of this effort, with based on the nuclear effects of parton distribution, we focus on the energy loss effect of the propagating quark in the cold nuclear medium. Baier, et al.(BDMPS)[41, 42] have computed perturbatively the radiated gluon spectrum by the hard quark in QCD media. By means of the BDMPS gluon spectrum, the quenching weight $P(\Delta E,\omega_{c},L)$ can be determined[43]. The analytic parameterization[33] on BDMPS quenching weight was given,
\begin{equation}
\omega_c P(\bar{\Delta E}=\Delta E/\omega_c) = \frac{1}{\sqrt{2\,\pi}\, \sigma\,\bar{\Delta E}}\,\exp\left[-\frac{\left(\log{\bar{\Delta E}}-\mu\right)^2}{2\,\sigma^2}\right],
\end{equation}
with $\mu=-2.55$ and $\sigma=0.57$ for the incoming quark energy $E$ independence.

We use the BDMPS quenching weight with the mean path length of the incoming quark in the nuclear medium $L=(3/4)R_A$. The quark transport coefficient $\hat{q}$ is treated as a constant $\hat{q}_0$ without any its kinematic variable dependence, which is below labelled the constant model. By taking advantage of minimizing $\chi^2$, the quark transport coefficient $\hat{q}$ is extracted with the parabolic error of the fit parameter from the MINUIT package[44]. The determined transport coefficient $\hat{q}$, the relative uncertainty and $\chi^2$ per number of degrees of freedom show that for the E906 measurements(12 data points), $\hat{q}_0=0.455\pm0.047$ GeV$^2$/fm, the relative uncertainty $\delta \hat{q}_0/\hat{q}_0\simeq10\%$ and $\chi^2/{\rm ndf}=0.889$. Fitting the E866 experimental data(16 data points) gives $\hat{q}_0=0.892\pm0.164$ GeV$^2$/fm with $\delta \hat{q}_0/\hat{q}_0\simeq18\%$ and $\chi^2/{\rm ndf}=0.362$. The computed Drell-Yan cross section ratios(blue dashed lines in Fig.1) are in agreement with E906 and E866 experimental data.

Now let us investigate the nuclear geometry effect on the Drell-Yan reaction of the proton off nuclei. In this case, the incoming colored quark will travel the path length $L=\sqrt{R_{\rm A}^2-b^2}+y$. By combining (4) with (8), the quark transport coefficient $\hat{q}$ and $\chi^2$ per number of degrees of freedom are calculated. It is found that for E906 experimental data, $\hat{q}_0=0.415\pm0.053$ GeV$^2$/fm with $\delta\hat{q}_0/\hat{q}_0\simeq13\%$ and $\chi^2/{\rm ndf}=0.644$. Regarding E866 experimental data, $\hat{q}_0=0.651\pm0.124$ GeV$^2$/fm with $\delta \hat{q}_0/\hat{q}_0\simeq19\%$ and $\chi^2/{\rm ndf}=0.370$. Our theoretical results(black solid lines in Fig.1) are consistent with the E906 and E866 experimental data. It is obvious that the nuclear geometry effect could reduce the quark transport coefficient by as much as $8.8\%$ for the E906 experiment and $27\%$ for E866 measurement, respectively. In addition, it is worth stressing that the different values of quark transport coefficient $\hat{q}_0$ are obtained from E906 and E866 experimental data. We think that the extracted different quark transport coefficients result from the different kinematic coverage by E906 and E866 experiments. As for E906 experiment, the Drell-Yan events were measured with the target-quark momentum fraction over the range $0.1<x_2<0.3$. However, the Drell-Yan dimuon production from E866 experiment was reported in the region $0.01<x_2<0.12$. It seems to say that the quark transport coefficient has the obvious dependence on the kinematic variable, i.e. the momentum fraction of a bound quark in the target nucleus. Unfortunately, after we try to carry out the global fit of E906 and E866 experimental data, our calculated result(green dot-dashed lines in Fig.1 and black solid lines in Fig.2) shows that $\hat{q}_0=0.455\pm0.052$ GeV$^2$/fm with $\delta\hat{q}_0/\hat{q}_0\simeq11\%$ and $\chi^2/{\rm ndf}=0.630$. A good description can still be given of the employed Drell-Yan data. Therefore, if we can confirm the conclusion that the quark transport coefficient depends on the target-quark momentum fraction, more accurate experimental data will be indispensable on the Drell-Yan differential cross section ratio.

In order to understand the detail of the quark transport coefficient, we need to pay attention to the gluon distribution function $G(x_{\rm G},Q^2_{\rm G})$ in the expression of $\hat{q}$. Following Ref.[16], if a fast quark of momentum $p$ traversing the target nucleus interacts with a bound nucleon of momentum $P$ by the exchanged gluon of momentum $q$, the Bjorken variable $x_{\rm G}$ in the gluon distribution is written as
\begin{equation}
x_{\rm G}\equiv\frac{q^{+}}{P^{+}}=\frac{q^{+}}{M_{\rm N}},
\end{equation}
in the target nucleus rest frame with using light-cone variables. In the above equation, $M_{\rm N}$ is the nucleon mass. In the proton-nucleus Drell-Yan reaction with a fixed target, the invariant mass of produced lepton pair $M^2=x_2 x_1s$, and $s\approx 2E_{\rm p}M_{\rm N}$. The momentum fraction $x_2$ of the quark from the bound nucleon
\begin{equation}
x_2= \frac{M^{2}}{2x_1E_{\rm p}M_{\rm N}},
\end{equation}
where $x_1E_{\rm p}$ is the energy of the incoming quark. Obviously, the Bjorken variable
\begin{equation}
x_{\rm G}=\frac{2x_1E_{\rm p} q^{+}}{M^2}x_2.
\end{equation}
Consequently, the unobservable quantity $x_{\rm G}$ in the quark transport coefficient $\hat{q}$ can be expressed with the measurable momentum fraction $x_2$, which is a Bjorken variable in nature.

As for the gluon density $x_{\rm G}G(x_{\rm G},Q^2_{\rm G})$ in the quark transport coefficient $\hat{q}$, the Bjorken variable $x_{\rm G}\ll1$, and the virtuality $Q^2_{\rm G}<1$ GeV$^{2}$. The gluon density $x_{\rm G}G(x_{\rm G},Q^2_{\rm G})$ can only be described by phenomenological non-perturbative model because perturbative QCD fails in the low virtuality $Q^2_{\rm G}$ and small Bjorken variable $x_{\rm G}$ region. In order to give a good description of the experimental data from deep inelastic scattering, Golec-Biernat and W\"{u}sthoff[45, 46] proposed a model based on the concept of saturation for small $Q^2_{\rm G}$ and small $x_{\rm G}$. According to the saturation model, the gluon density
\begin{equation}
x_{\rm G}G(x_{\rm G},Q^2_{\rm G})\sim x_{\rm G}^{-\lambda}.
\end{equation}
After  utilizing the Bjorken variable $x_2$ instead of $x_{\rm G}$,
\begin{equation}
x_{\rm G}G(x_{\rm G},Q^2_{\rm G})\sim x_2^{-\lambda}.
\end{equation}
Therefore, we can introduce Bjorken variable $x_2$ dependence of the quark transport coefficient by writing
\begin{equation}
\hat{q}=\hat{q_0}x_2^{\alpha},
\end{equation}
which is referred to as the power-law model hereafter. Two undetermined parameters with the power-law model are $\hat{q_0}$ and $\alpha$.

For the power-law model, a global analysis on all 28 data points from E906 and E866 measurement gives that $\hat{q}_0=0.247\pm0.027$ GeV$^2$/fm with the relative uncertainty $\delta \hat{q}_0/\hat{q}_0\simeq11\%$, $\alpha=-0.280\pm0.043$ with $\delta\alpha/\alpha\simeq15\%$ and $\chi^2/{\rm ndf}=0.484$. The rather small value of $\chi^2/{\rm ndf}$ indicates a good fit quality, which is also seen from our calculated results(red dashed lines in Fig.2) compared with E906 and E866 experimental data. As shown by the black solid line in Fig.3, the quark transport coefficient $\hat{q}$ decreases with the increasing Bjorken variable $x_2$ in the power-law model.

\begin{figure}
\centering
\includegraphics*[width=15cm,height=10.7cm]{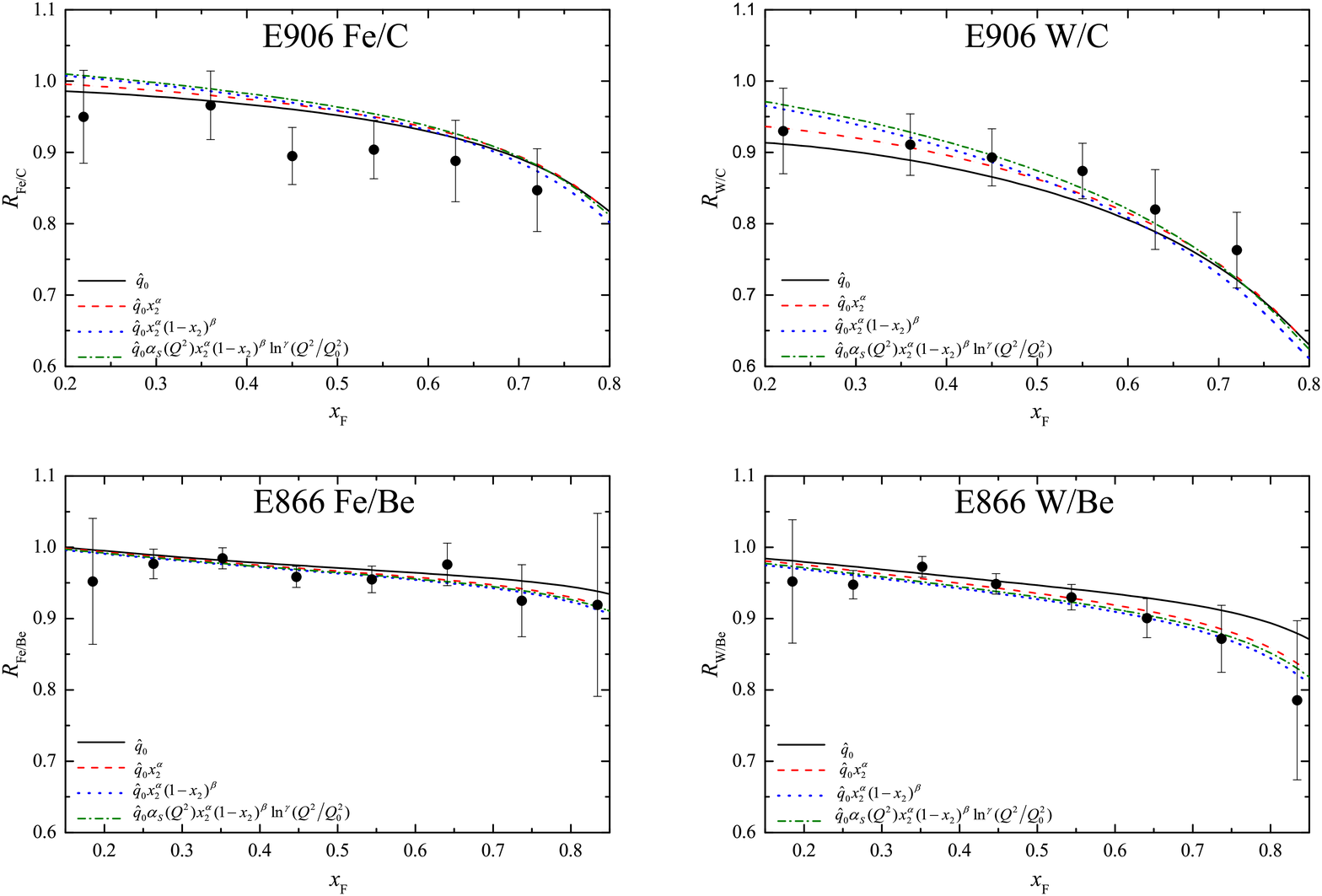}
\vspace{-0.1cm}
\caption{The calculated Drell-Yan cross section ratios $R_{\rm A/B}$ as a function of $x_{\rm F}$ from the parameterization of BDMPS quenching weight for the quark energy $E$ independence. The different lines correspond respectively to the results from $\hat{q}$ in four different models. The experimental datas are shown with the total uncertainty (statistical plus systematic, added quadratically).}
\end{figure}

Regarding the gluon density $x_{\rm G}G(x_{\rm G},Q^2_{\rm G})$, after the unmeasurable $x_{\rm G}$ is replaced with the observable Bjorken variable $x_2$, we should consider the intermediate and large Bjorken variable $x_2$ correction to the quark transport coefficient. Hence, we can suppose Bjorken variable $x_2$ dependence of the quark transport coefficient,
\begin{equation}
\hat{q}=\hat{q_0}x_2^{\alpha}(1-x_2)^{\beta},
\end{equation}
which is hereafter referred to as the double power-law model. Three parameters of the fits in the double power-law model are $\hat{q_0}$, $\alpha$ and $\beta$.

In the double power-law model, the global fit to the experimental data shows that $\hat{q}_0=0.684\pm0.072$ GeV$^2$/fm with the relative uncertainty $\delta \hat{q}_0/\hat{q}_0\simeq11\%$, $\alpha=-0.081\pm0.040$ with $\delta\alpha/\alpha\simeq49\%$, $\beta=3.825\pm0.775$ with $\delta\beta/\beta\simeq20\%$ and $\chi^2/{\rm ndf}=0.602$. The numeric results of the Drell-Yan cross section ratios correspond to the blue dotted lines in Fig.2. By taking into account the data precision, the theoretical results are in good agreement with the experimental data. The trend of $\hat{q}$ with respect to $x_2$ has been drew as a red dashed line in Fig.3. It can be found that in the region $x_2>0.156$, $\hat{q}$ decreases rapidly with the increasing $x_2$.

\begin{figure}
\centering
\includegraphics*[width=9cm,height=7cm]{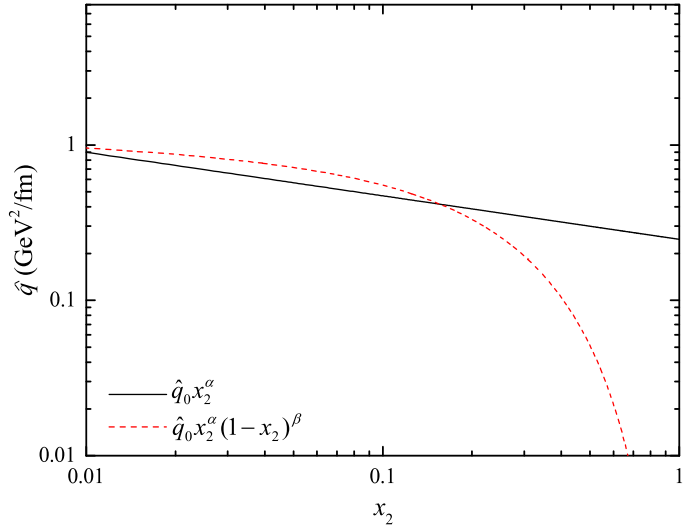}
\vspace{-0.1cm}
\caption{The obtained quark transport coefficient $\hat{q}$ as a function of Bjorken variable $x_2$ from power-law model and double power-law model, respectively.}
\end{figure}

Although the power-law model and the double power-law model provides good quality fit to the experimental data from E906 and E866, these two models do not reflect the scale $Q^2$ dependence of $\hat{q}$. Therefore, both power-law model and double power-law model can not provide full information about the quark transport coefficient $\hat{q}$.

Now that $x_{\rm G}$ can be substituted by Bjorken variable $x_2$, the virtuality $Q^2_{\rm G}$ should also be expressed with the measurable scale $Q^2$ though the detail is not clear. Then, the strong coupling constant $\alpha_{\rm s}(Q^2_{\rm G})$ in quark transport coefficient $\hat{q}$ can be written as $\alpha_{\rm s}(Q^2)$. Along with the evolution of gluon distribution with $Q^2$, the quark transport coefficient is assumed to be
\begin{equation}
\hat{q}=\hat{q_0} \alpha_{\rm s}(Q^2) x_2^{\alpha}(1-x_2)^{\beta}\ln^{\gamma}(Q^2/Q^2_0),
\end{equation}
which is hereafter known as the evolution model with $Q^2_0 = 1$ GeV$^2$ in order to ensure the argument in the logarithm dimensionless. The parametrization form is the same as that in Ref.[25].
Four parameters with the evolution model are $\hat{q_0}$, $\alpha$, $\beta$ and $\gamma$.

\begin{figure}
\centering
\includegraphics*[width=9cm,height=7cm]{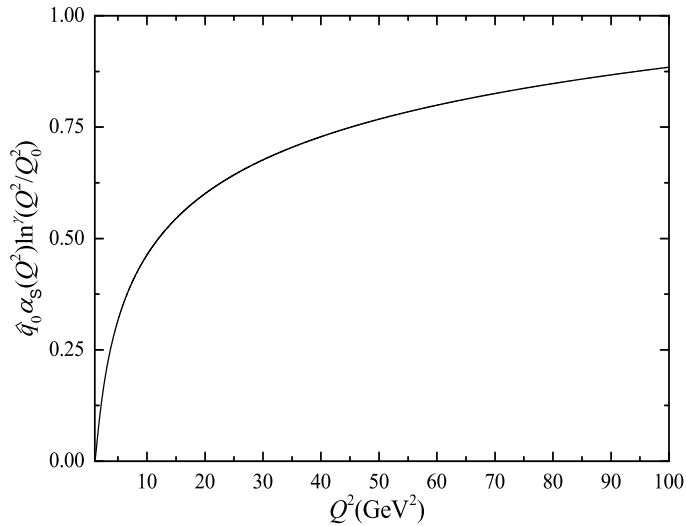}
\vspace{-0.1cm}
\caption{The part $\hat{q}_0\alpha_{\rm s}\ln^\gamma(Q^2/Q_0^2)$ in $\hat{q}$ varying with $Q^2$ in the evolution model.}
\end{figure}

We perform the global fit of $\hat{q}$ in the evolution model and extract the parameter values. It is shown that $\hat{q}_0=0.553\pm0.058$ GeV$^2$/fm with the relative uncertainty $\delta \hat{q}_0/\hat{q}_0\simeq10\%$, $\alpha=-0.080\pm0.040$ with $\delta\alpha/\alpha\simeq50\%$, $\beta=3.796\pm0.773$ with $\delta\beta/\beta\simeq20\%$, and $\gamma=1.436\pm0.085$ with $\delta\gamma/\gamma\simeq6\%$ and $\chi^2/{\rm ndf}=0.609$. The calculated Drell-Yan cross section ratios(green dot-dashed lines in Fig.2) agree with the E906 and E866 experimental data. Fig.4 gives the $Q^2$ part of $\hat{q}$ varying with $Q^2$ variable. It is found that the factor $\hat{q}_0\alpha_{\rm s}\ln^\gamma(Q^2/Q_0^2)$ increases with $Q^2$ rapidly in the range $Q^2<20{\rm GeV}^2$ and slowly in $Q^2>20{\rm GeV}^2$. In order to see the panorama of quark transport coefficient, the determined $\hat{q}$ as a function of $x_2$ and $Q^2$ is shown in Fig.5 for the evolution model.

\begin{figure}
\centering
\includegraphics*[width=15cm,height=11.5cm]{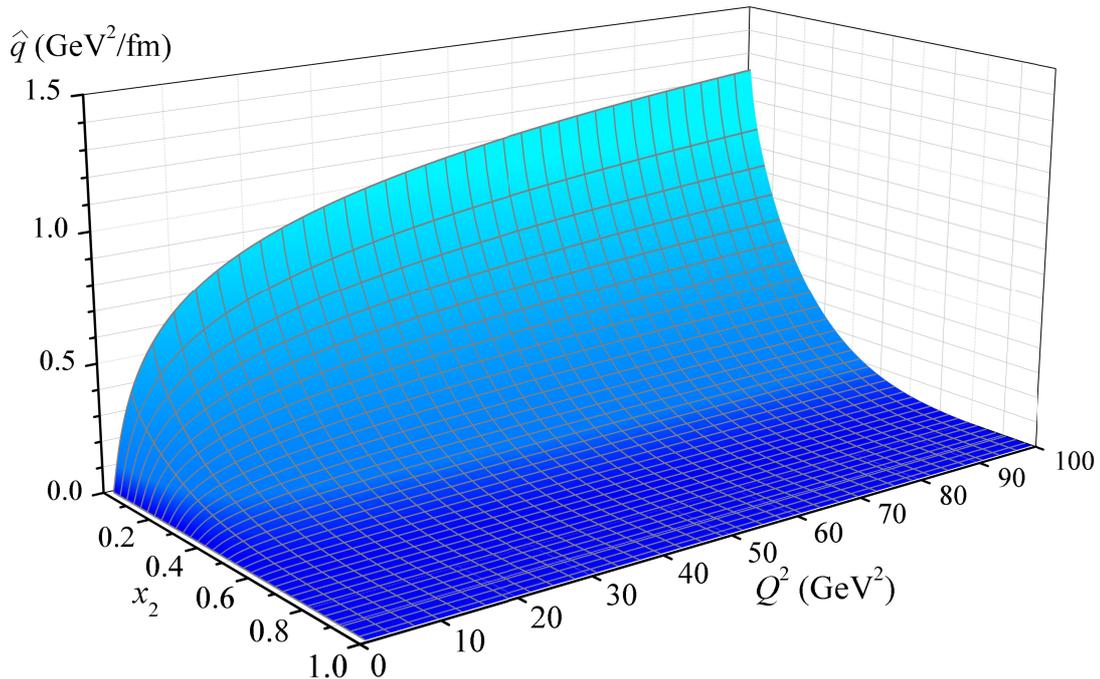}
\vspace{-0.1cm}
\caption{The obtained quark transport coefficient $\hat{q}$ as a function of Bjorken variable $x_2$ and scale $Q^2$.}
\end{figure}

In Table.I, we summarize the obtained parameter values of the quark transport coefficient from our fits in above four models: the constant model, the power-law model, the double power-law model and the evolution model. It is found that the power $\alpha$ in the factor $x^{\alpha}$ has a big discrepancy between the power-law model and double power-law model because of the impact from the middle and large $x_2$. A narrow different of the value of $\alpha$ and $\beta$ between the evolution model and double power-law model indicates that the $x_2$ part of $\hat{q}$ varying with $x_2$ in the evolution model follows the same trend as that from double power-law model. It should be point out that the relative uncertainty of parameter $\alpha$ in the double power-law model and evolution model is very large. The improvement is needed in the future study.

\begin{table}
\caption{The parameter values of $\hat{q}$ and $\chi^{2}/{\rm ndf}$ extracted in four different models from E866 and E906 experimental data.}
\begin{ruledtabular}
\begin{tabular*}{\hsize}
{c@{\extracolsep{0ptplus1fil}} c@{\extracolsep{0ptplus1fil}}
c@{\extracolsep{0ptplus1fil}} c@{\extracolsep{0ptplus1fil}}
c@{\extracolsep{0ptplus1fil}} c@{\extracolsep{0ptplus1fil}}}

  $\hat{q}$(GeV$^2$/fm)                                                               &$\hat{q_0}$    &$\alpha$         &$\beta$         &$\gamma$        &$\chi^{2}/{\rm ndf}$\\
  \hline
  $\hat{q_0 }$                                                                        &$0.455\pm0.052$ &$                $& $           $&$ $             &$0.630$\\
  $\hat{q_0}x^{\alpha}$                                                               &$0.247\pm0.027$ &$-0.280\pm0.043$ &$             $&$ $             &$0.484$\\
  $\hat{q_0}x^{\alpha}(1-x)^{\beta}$                                                  &$0.684\pm0.072$ &$-0.081\pm0.040$ &$3.825\pm0.775$&$ $             &$0.602$\\
  $\hat{q_0}\alpha_{\rm s}(Q^{2})x^{\alpha}(1-x)^{\beta}\ln^{\gamma}(Q^{2}/Q^{2}_0)$  &$0.553\pm0.058$ &$-0.080\pm0.040$ &$3.796\pm0.773$&$1.436\pm0.085$ &$0.609$\\

\end{tabular*}
\end{ruledtabular}
\end{table}

It is worth mentioning that Fran\c{c}ois Arleo et al.[14, 16] studied the transport coefficient parameterization as a function of Bjorken variable. Their research did not include the virtuality $Q^2$ evolution of transport coefficient. In addition,we adopte the same functional form to parameterize jet transport coefficient as that by Ru et al.[25]. However, the extracted parameter values of jet transport coefficient are different from those given by our global analysis on Drell-Yan experimental data. The reason remains to be further explored in the future.

\section{Summary}
In this paper we study the measurable kinematic variables dependence of quark transport coefficient in nuclear Drell-Yan process. By means of the nuclear parton distributions determined only with lepton-nuclear deep inelastic scattering experimental data and the analytic parameterization of quenching weight based on BDMPS formalism, the leading order calculations of the nuclear Drell-Yan differential cross section ratio as a function of Feynman variable are carried out and compared with the experimental data from Fermilab E906 and E866 Collaborations. It is found that only the nuclear effects of parton distribution can not explain the employed experimental data. Apart from the nuclear effects on the parton distribution, there is the incoming quark energy loss effect in nuclear Drell-Yan process. In a condition of the average path length of the incident quark in the nucleus and the quark transport coefficient as a constant, the calculated results give that the values of quark transport coefficient $\hat{q}$ are respectively $0.455\pm0.047$ GeV$^2$/fm for E906 measurement and $0.892\pm0.164$ GeV$^2$/fm for E866 experimental data. In view of the nuclear geometry effect on the nuclear Drell-Yan reaction, our calculations show that the quark transport coefficients are $0.415\pm0.053 $ GeV$^2$/fm and $0.651\pm0.124$ GeV$^2$/fm for E906 and E866 measurements, respectively. It is evident that the nuclear geometry effect has a significant impact on the quark transport coefficient in cold nuclear matter. Therefore, we should consider the nuclear geometry effect in studying the nuclear Drell-Yan process. However, if we want to confirm the conclusion that the quark transport coefficient depends on the target-quark momentum fraction, more accurate experimental data will be necessary on the Drell-Yan differential cross section ratio. Fortunately, it is discovered that there exists the relation between the unobservable Bjorken variable in the quark transport coefficient and the measurable kinematic variable in Drell-Yan process. Furthermore, three models are discussed with the quark transport coefficient as a function of the measurable kinematic variables. Although the power-law model and the double power-law model all display high quality fit, these two models can not reflect the complete information on quark transport coefficient. As for the so-called evolution model, the quark transport coefficient is determined as a function of the Bjorken variable $x_2$ and scale $Q^2$.

Exploring the depths of the parton propagation in cold nuclear matter is a better benchmark for the physical mechanisms by which the traversing partons lose energy to, and interact with the quark-gluon plasma. The Drell-Yan reaction in hadron-nucleus collisions provides an excellent place. In order to gain precise knowledge about the quark transport coefficient in a cold nuclear medium, further theoretical and experimental researches[47] are needed on the basis of the existing researches[16, 24-26, 32]. The parametrization form of the quark transport coefficient can be further optimized for precise understanding of the parton propagation mechanisms in nuclear matter.

\vskip 1cm
{\bf Acknowledgments}
We thank Professor Zhi-Hui Guo for interesting and useful discussions. This work is supported in part by the National Natural Science Foundation of China(11575052, 11975090).

\end{document}